\documentstyle[prc,graphicx,floats,aps]{revtex}

\begin{document}

\draft
\twocolumn[\columnwidth\textwidth\csname@twocolumnfalse\endcsname
\title{Neutrino spectra from stellar electron capture}

\author{K. Langanke$^1$, G. Mart\'{\i}nez-Pinedo$^{1,2}$ and J. M.
  Sampaio$^1$}
 
\address{$^1$Institut for Fysik og Astronomi, {\AA}rhus Universitet,
  DK-8000 {\AA}rhus C, Denmark\\
  $^2$ Departement f\"ur Physik und Astronomie, Universit\"at Basel,
  Basel, Switzerland }

\date{\today}
\maketitle

\begin{abstract}
  Using the recent shell model evaluation of stellar weak interaction
  rates we have calculated the neutrino spectra arising from electron
  capture on $pf$-shell nuclei under presupernova conditions. We
  present a simple parametrization of the spectra which allows for an
  easy implementation into collapse simulations. We discuss that the
  explicit consideration of thermal ensembles in the parent nucleus
  broadens the neutrino spectra and results in larger average neutrino
  energies. The capture rates and neutrino spectra can be easily
  modified to account for phase space blocking by neutrinos which
  becomes increasingly important during the final stellar collapse.
\end{abstract}

]

Weak interactions play a central role in the final evolution of
massive stars \cite{Woosley84}.  These processes, mainly electron
capture and beta decays, create neutrinos which for all densities
until the collapse becomes truly hydrodynamic (i.e. $\rho \sim
10^{11}$ g cm$^{-3}$) escape the star carrying away energy and
reducing the entropy. Thus, in simulations which follow the star's
evolution until the iron core reaches central densities of order a few
$10^9$ g cm$^{-3}$ it is sufficient to know the average neutrino
energies of the various weak processes to determine the energy loss
rate along the stellar trajectory, e.g. \cite{Heger00}.  Such
simulations define the `presupernova models', e.g. \cite{WW95}, which
are then being used as input to detailed studies of the collapse and
explosion mechanism, e.g.  \cite{Mezzacappa00}. In recent years it has
become apparent that the various neutrino reactions play an essential
role in these simulations and, as neutrino reactions depend
sensitively on energy, these studies require a detailed bookkeeping of
the neutrino spectra. This goal is achieved within hydrodynamical
models with explicit neutrino Boltzmann transport
\cite{Bruenn85,Mezzacappa93,Janka96}.  The necessary input into such
simulations are the weak rates and the associated neutrino spectra.

So far, electron capture on nuclei has only been treated quite
schematically in collapse simulations with neutrino Boltzmann
transport, e.g. \cite{Bruenn85,Mezzacappa93,Janka96}. The ensemble of
nuclei in the stellar composition, given by nuclear statistical
equilibrium, is represented by an `average nucleus', whose capture
rate then is derived on the basis of the independent particle model
(even reduced to a model which only considers $f_{7/2}$ and $f_{5/2}$
orbitals \cite{Bruenn85}). The corresponding neutrino spectrum is then
approximated in the spirit of this model, assuming a
nucleus-independent energy splitting of 3 MeV between the $f_{7/2}$
and $f_{5/2}$ orbitals.  We note, that the standard for the stellar
weak interaction rates in presupernova evolutions, however, was set by
the work of Fuller, Fowler and Newman (FFN) \cite{FFN1,FFN2} who
estimated the rates for the nuclei with mass numbers $A=21-60$ on the
basis of the independent particle model and experimental data,
whenever available. Presupernova evolution studies then considered the
FFN rate tables for a proper nuclear composition. FFN list the average
neutrino energies for the various weak reactions, but to our knowledge
neutrino spectra have not been derived on the basis of the FFN
compilation. As mentioned above, such spectra are also not required
for presupernova studies as the neutrinos leave the star unhindered.

Due to progress in nuclear many-body modeling and in computer hardware
and guided by experimental data it has recently been possible to treat
the nuclear structure problem involved in the calculation of stellar
weak interaction rates in a reliable way \cite{Langanke00,ADNDT}.  The
calculations have been performed on the basis of large-scale shell
model studies which reproduce all experimentally available relevant
data quite accurately \cite{Caurier99}.  These shell model rates show
some marked differences to the FFN estimates, leading to significant
changes in the presupernova evolution of massive stars
\cite{Heger00,HegerPRL}.  It appears therefore reasonable that this
compilation \cite{ADNDT} should also be used in collapse and explosion
studies which build on the presupernova models. To make such use
possible we will here study the neutrino spectra corresponding to the
shell model rates and show a way how these spectra can be easily and
consistently implemented in collapse codes. It is important to note
that under the conditions of the presupernova models and in the
subsequent stellar evolution beta decay is strongly blocked by the
appreciable electron chemical potential and the total electron capture
rates are orders of magnitude larger
\cite{Langanke00,Heger00,HegerPRL}. Thus it is quite sufficient to
focus on the neutrino spectra arising from electron captures for
post-presupernova simulations where detailed neutrino transport is
important.

Under the stellar conditions we are concerned with 
electron capture is dominated  by Gamow-Teller (GT) transitions
\cite{BBAL}. The appropriate formalism has been derived in
\cite{FFN1,FFN2}:

\begin{eqnarray}
  \label{eq:rate}
  \lambda &=& \frac{\ln 2}{K} 
  \left(\frac{g_A}{g_V}\right)^2
\sum_{i,j} \frac{(2J_i+1)
    e^{-E_i/(kT)}}{G(Z,A,T)}
  \frac{|\langle j||\sum_k {\sigma}^k {t}^k_+ || i
  \rangle|^2}{2 J_i +1} \nonumber \\
  & \times &  \int_{w_l}^\infty w  p  (Q_{ij}+w)^2
    F(Z,w)  
       S_e(w) dw,
\end{eqnarray}
where the sums in $i$ and $j$ run over states in the parent and
daughter nuclei, respectively.  For the constant $K$ we used
$K=6146\pm 6$~s~\cite{Towner} and $g_V, g_A$ are the vector and
axialvector coupling constants.  $G(Z,A,T)=\sum_i \exp(-E_i/(kT))$ is
the partition function of the parent nucleus.  The sum in the GT
matrix element runs over all nucleons.  In the phase space integral
$w$ is the total energy of the electron in units of $m_e c^2$, and
$p=\sqrt{w^2-1}$ is its momentum in units of $m_e c$.  Finally the
Q-value for a transition between two nuclear states $i,f$ is defined
in units of $m_e c^2$ as
\begin{equation}
  \label{eq:qn}
    Q_{ij} = \frac{1}{m_e c^2} (M_p - M_d + E_i -E_j),
\end{equation}
where $M_p, M_d$ are the nuclear masses of the parent and daughter
nucleus, respectively, while $E_i, E_j$ are the excitation energies of
the initial and final states.  The lower integral limit in (1) is
$w_l=1$ if $Q_{ij} > -1$ or $w_l = |Q_{ij}|$ if $Q_{ij} < -1$.  For
the stellar conditions we are interested in, the electrons are well
described by Fermi-Dirac (FD) distributions, with temperature $T$ and
chemical potential $\mu_e$:
\begin{equation}
  \label{eq:fermie}
  S_e (E_e) = \frac{1}{\exp\left(\frac{E_e - \mu_e}{kT}\right)+1},
\end{equation}
with $E_e = w m_e c^2$.  The remaining factor appearing in the phase
space integrals is the Fermi function, $F(Z,w)$, that corrects the
phase space integral for the Coulomb distortion of the electron wave
function near the nucleus.

Applying the formalism described above and determining the nuclear
matrix elements within large-scale shell model calculations, stellar
electron capture rates have recently been calculated for $pf$-shell
nuclei. These nuclei dominate the weak processes in the presupernova
evolution of massive stars and details of the rate evaluations can be
found in \cite{Langanke00}. Here we will use the same approach to
study the neutrino spectra emerging in the electron capture reactions.
Energy conservation requires that the neutrino emitted after capture
of an electron with energy $w$ on an initial state $i$ leading to a
final state $j$ is:
\begin{equation}
E_\nu = m_e c^2 (w +Q_{ij}) .
\end{equation}

The neutrino spectra for a specific nuclear transition $i \rightarrow
j$ is given by the respective partial rate per energy interval. The
total spectrum is then the sum over all possible transitions. The
normalized neutrino spectrum $n(E_\nu)$ is obtained by dividing by the
total electron capture rate.

In the derivation above we have explicitly assumed that the neutrinos
produced by the electron capture process leave the star; i.e. there is
no blocking of the phase space due to the presence of neutrinos in the
stellar environment. The spectra, which we will discuss below, are all
derived based on this assumption. In the collapse phase following the
presupernova evolution neutrinos are getting increasingly trapped in
the core. This will require the inclusion of a neutrino blocking
factor $(1-S_\nu(E_\nu))$ in the phase space integral of the electron
capture formula, Eq. (1). The neutrino distribution $S_\nu (E_\nu)$
depends on position and time and can be calculated within the
Boltzmann transport formalism. Once $S_\nu (E_\nu)$ is known, the
present neutrino spectra and the capture rates of \cite{Langanke00}
can be easily corrected for neutrino blocking. For the neutrino
spectra this is achieved by folding the uncorrected spectra
$n(E_\nu)$, as presented in this paper, with the blocking factor
$(1-S_\nu(E_\nu))$. The corrected capture rates are obtained by
multiplying the tabulated rates \cite{ADNDT} with $\int n(E) (1-S_\nu
(E)) dE$.

For the following discussion it is useful to realize that the neutrino
spectra depend basically on three quantities
\begin{itemize}
\item
the electron chemical potential
\item the `effective' Q-value, $Q_{\text{eff}}=M_p-M_d+E_i$
\item
the GT strength distribution.
\end{itemize}

Generally one expects that neutrino energies are increased for larger
electron chemical potentials, favorable mass differences between
daughter and parent nuclei $(M_p-M_d)$, and from thermally excited
states. Furthermore, large neutrino energies are favored if the strong
GT transition resides at low excitation energies in the daughter
nucleus. Due to nuclear pairing structure arguments \cite{Langanke00}
this is the case for odd-odd daughter nuclei (capture on even-even
parents), while the bulk of the GT strength is somewhat higher ($\sim
2- 3$ MeV) for odd-A nuclei and is shifted by additional 2-3 MeV in
even-even daughter (capture on odd-odd parents) \cite{Caurier99}.  We
note that these differences in the GT distributions result in the fact
that low-lying rather weak GT transitions contribute relatively more
to the electron capture rates on odd-odd nuclei and odd-A nuclei,
while for even-even parents the GT bulk often resides at such low
energies (e.g. Ni isotopes) that a distiction between low-lying
strength and GT bulk is not very meaningful.

To illustrate the discussion we plot in Fig. 1 the normalized neutrino
spectra for electron capture on the A=56 isobars. The calculation has
been performed for typical conditions during silicon shell burning of
a 15~M$_\odot$ star \cite{Heger00} ($T=4 \times 10^9$ K, $\rho =3
\times 10^8$ g cm$^{-3}$, $Y_e=0.45$).  The resulting electron
chemical potential then is $\mu_e=2.5$ MeV.

\begin{figure}[thb]
  \begin{center}
     \includegraphics[width=0.9\columnwidth]{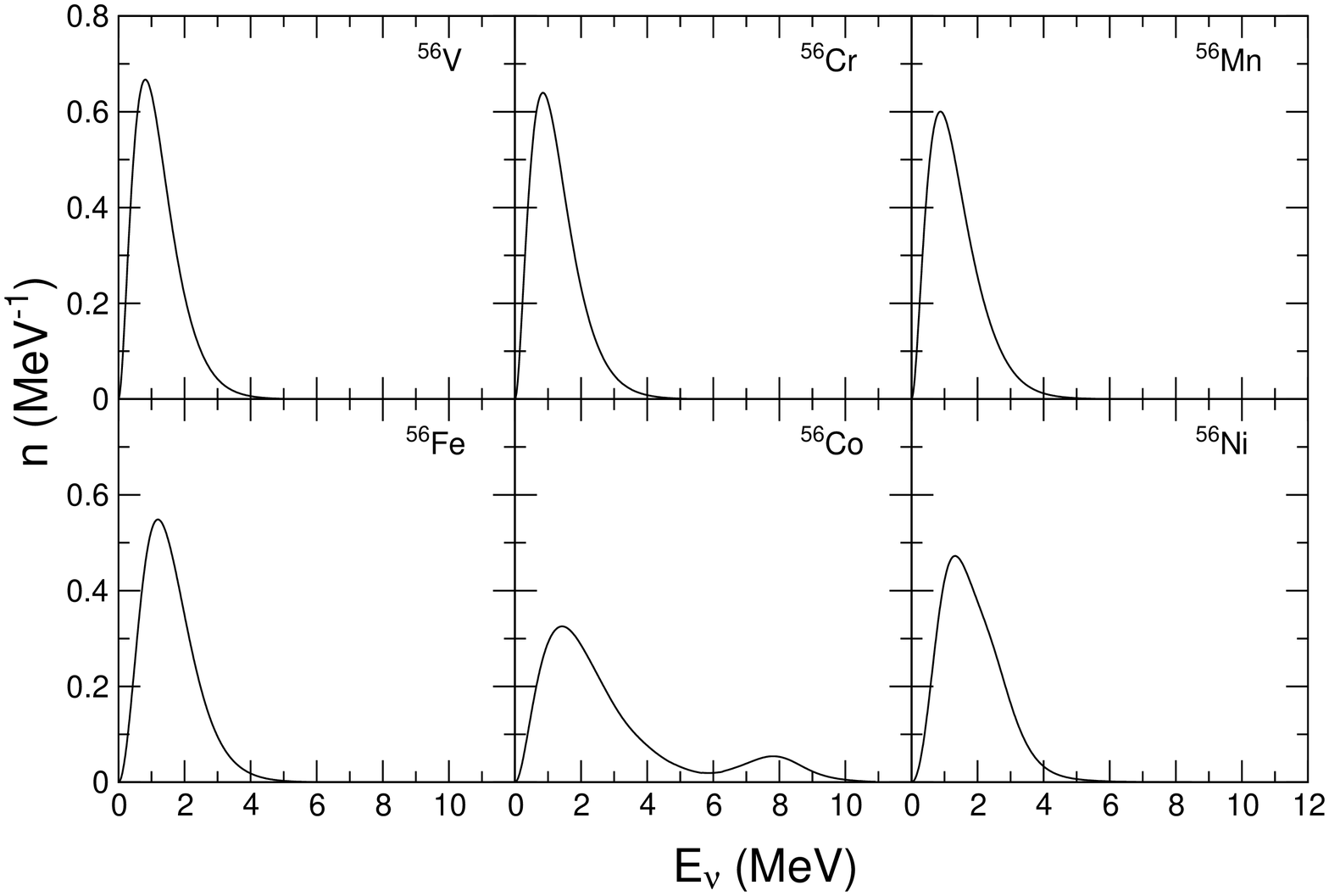}
  \end{center}
  \caption{Normalized neutrino spectra for stellar electron capture on
    selected $A=56$ isobars. The spectra have been calculated for
    stellar conditions which are typical for silicon shell burning in
    a 15~M$_\odot$ star \protect\cite{Heger00}.}
\end{figure}

We note that $^{56}$V, $^{56}$Cr, $^{56}$Mn and $^{56}$Fe have
$Q_0=M_p-M_d <0$, thus making electron capture in the laboratory
impossible. The other two nuclei $^{56}$Co ($Q_0= 4.06$ MeV) and
$^{56}$Ni ($Q_0 = 1.62$ MeV) decay dominantly by electron capture.
With the exception of $^{56}$Co the neutrino spectra for the other 5
nuclei are very similar: they are peaked around rather small neutrino
energies $E_\nu \approx 1-2$ MeV with a width of 1.4-1.8 MeV. The
reason for this quite similar structure is twofold. For the nuclei
with negative $Q_0$-values, electron capture is hindered and requires
electrons from the exponentially decreasing tail of the FD
distribution.  Obviously to achieve an appreciable rate it is
advantegeous to keep the neutrino energies small.  Thus capture to
low-lying states occurs with electrons with lower energies than
capture to the bulk of the GT strength, but both are accompanied by
low-energy neutrinos. For $^{56}$Ni the $Q_0$-value allows capture of
electrons with all energies.  However, the GT distribution in the
daughter $^{56}$Co is well concentrated at low excitation energies,
resulting again in a rather narrow neutrino spectrum.

\begin{figure}[thb]
  \begin{center}
     \includegraphics[width=0.9\columnwidth]{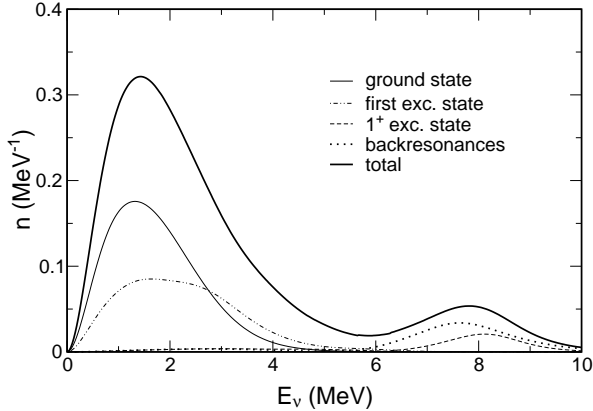}
  \end{center}
  \caption{Partial contributions of individual states in the parent
    nucleus to the neutrino spectrum for stellar electron capture on
    $^{56}$Co. The calculation has been performed for the same
    conditions as in Fig.  1.  The spectra are multiplied by their
    relative weight to the total capture rate.}
\end{figure}

The $^{56}$Co spectrum is quite different, showing a double-bump
structure.  The reason is explained in Fig. 2 which shows the partial
neutrino spectra contributed by selected states in the parent nucleus.
The spectra for both, the ground state ($J=4$) and the first excited
state ($J=3$) show the single-peak structure, as generally observed
for the other nuclei. We also find the spectra for the first excited
state somewhat wider than for the ground state. This reflects the
strong angular momentum mismatch for the ground state which does not
connect to states in $^{56}$Fe below 2 MeV and has only very weak
transitions to states below 3 MeV. This is different for the excited
state which can reach the lower-lying $J=2$ states in $^{56}$Fe and in
that way produces neutrinos with larger energies. Interestingly the
excited $J=1$ state at 1.7 MeV excitation energy in $^{56}$Co produces
a neutrino spectrum with a peak energy around $E_\nu=8$ MeV. We note
that this state has a rather strong GT matrix element to the $^{56}$Fe
ground state and hence, considering its favorable effective Q value of
$Q_{\text{eff}} \sim 5.8$ MeV, electron capture on this state
generates neutrinos with rather large energies. While the excitation
energy in the parent increases the effective Q-value, and thus the
average neutrino energy, it reduces the contribution to the rate due
to the Boltzmann weight.  However, the excited $1^+$ state yields the
clue to the higher-energy neutrino peak.  We notice that there are
many more states in the excitation energy range $\sim$~2--4~MeV which
are connected to the low-lying states in the daughter nucleus
$^{56}$Fe by strong GT transitions. FFN have coined the term
`backresonances' for these states \cite{FFN1,FFN2} as they are part of
the bulk of the GT strength built on the low-lying states in the
inverse direction (Ref.  \cite{Langanke00} explains how these states
are considered in the rate evaluation.) Electron capture on these
backresonances occurs with favorable $Q_{\text{eff}}$-value and hence
allows the emission of neutrinos with rather high energies. Despite
the Boltzmann suppression the gain in phase space combined with the
large matrix elements ensure that these states combined contribute
noticeably to the total rate and produce the second peak in the total
neutrino spectrum.  Noting that the integral over the spectrum
reflects the relative contribution to the total rate, we remark in
passing that Fig. 2 also implies that at the chosen conditions
electron capture on $^{56}$Co is dominated by the one on the ground
state.  Although similar in nuclear structure, the electron capture on
the odd-odd nuclei $^{56}$Mn and $^{56}$V does not produce a
double-bump structure due to the negative $Q_0$ value which favors
emission of low-energy neutrinos.

\begin{figure}[thb]
  \begin{center}
     \includegraphics[width=0.9\columnwidth]{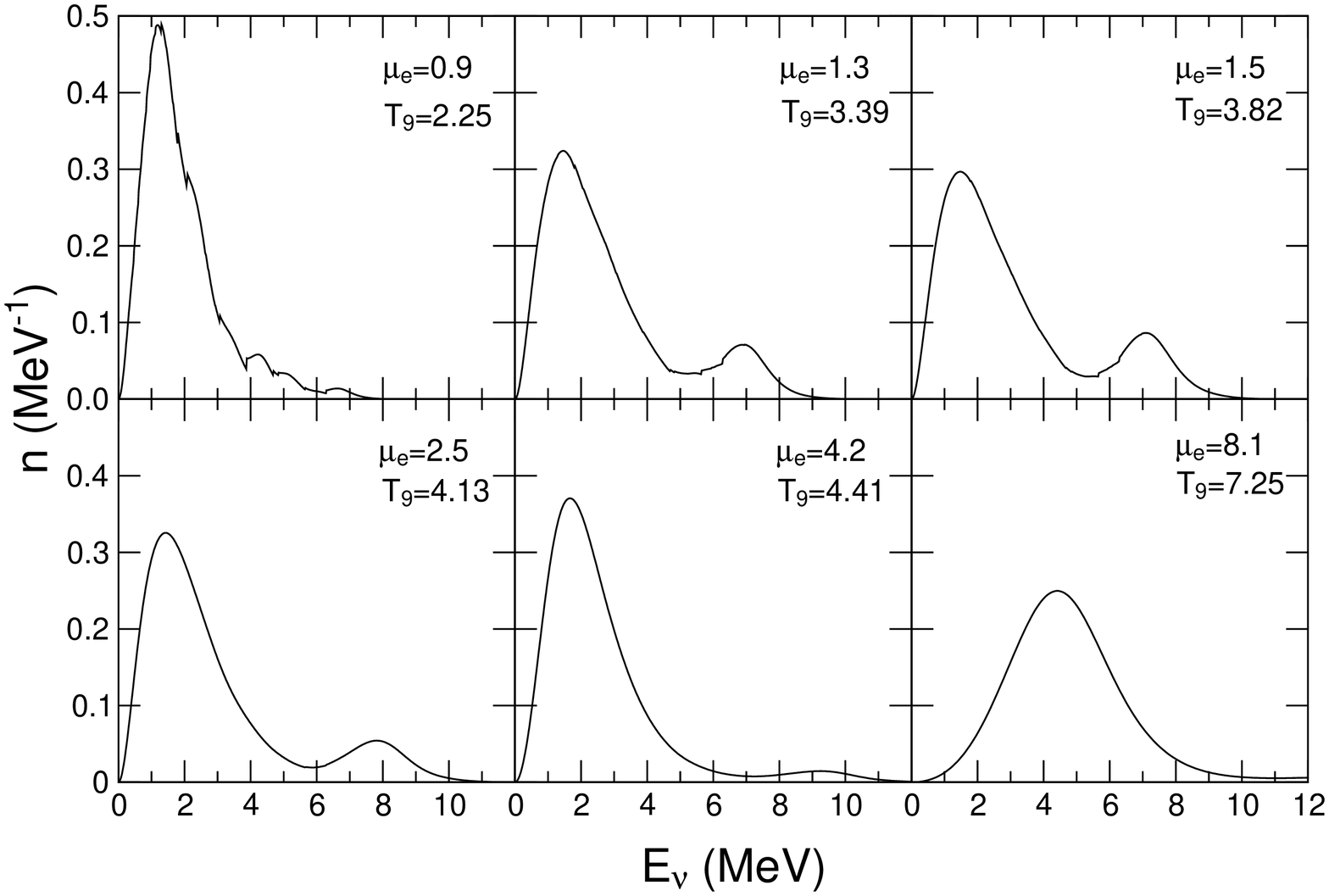}
  \end{center}
  \caption{Normalized neutrino spectra for stellar electron capture on 
    $^{56}$Co at several different phases of the final evolution of a
    15~$M_\odot$ star. The stellar parameters have been taken from
    Table~1 of \protect\cite{Heger00}. The chemical potentials are
    given in MeV, while $T_9$ defines the temperature in $10^9$ K.}
\end{figure}

At the conditions of silicon shell burning, depicted in Fig. 2, the
electron chemical potential is yet not large enough ($\mu_e=2.5$ MeV)
to allow significant capture on the odd-odd nucleus $^{56}$Co from
low-lying states to the bulk of the GT$_+$ distribution in the
daughter nucleus which resides at around $\sim$~7--9~MeV in
$^{56}$Fe~\cite{Martinez99}.  However, in the subsequent stellar
evolution $\mu_e$ increases rather fast.  Thus, capture to the bulk,
e.g. with significantly larger GT strength, becomes easier for the
low-lying states. This increases their relative weight compared with
the one of the backresonances and is also not compensated by the
relative gain in the Boltzmann factor of the latter. This behavior is
demonstrated in Fig. 3, again for $^{56}$Co.  The temperatures have
been chosen accordingly, using the stellar trajectories as given by
Heger {\em et al.} \cite{Heger00}.  We also observe that, once capture
to the bulk of the GT distribution dominates the capture rate, the
spectrum becomes single-peaked (approximately for $\mu_e=4.2$). A
further increase in $\mu_e$ then simply increases the average neutrino
energy (e.g. for $\mu_e=8.1$).

We summarize that the relative height of the two peaks reflects the
competition between electron chemical potential (its decrease reduces
the capture from low-lying states to the bulk of the GT distribution)
and temperature (its decrease reduces the Boltzmann weight of the
backresonances).  If one follows the stellar evolution backwards in
time, i.e.  to smaller temperatures, densities and electron chemical
potential, the spectrum ultimatively becomes single-peaked as it is
dominated by the temperature-favored ground state contribution.  The
spectrum also develops `discontinuities' (see e.g. for $\mu_e=0.9$)
which reflect the fact that the spectrum represents a sum over several
initial and final states which all have a definite minimal neutrino
energy $E_\nu^{\text{min}}=m_e c^2 (Q_{ij}+1)$.  At
$E_\nu^{\text{min}}$, the electron momentum vanishes ($p=0$), but $p
F(z,\omega)$ is finite resulting in a finite value for
$n(E_\nu^{\text{min}})$.  For larger $\mu_e$ values these
discontinuities are smeared out as the individual spectra noticeably
overlap.

\begin{figure}[thb]
  \begin{center}
    \includegraphics[width=0.9\columnwidth]{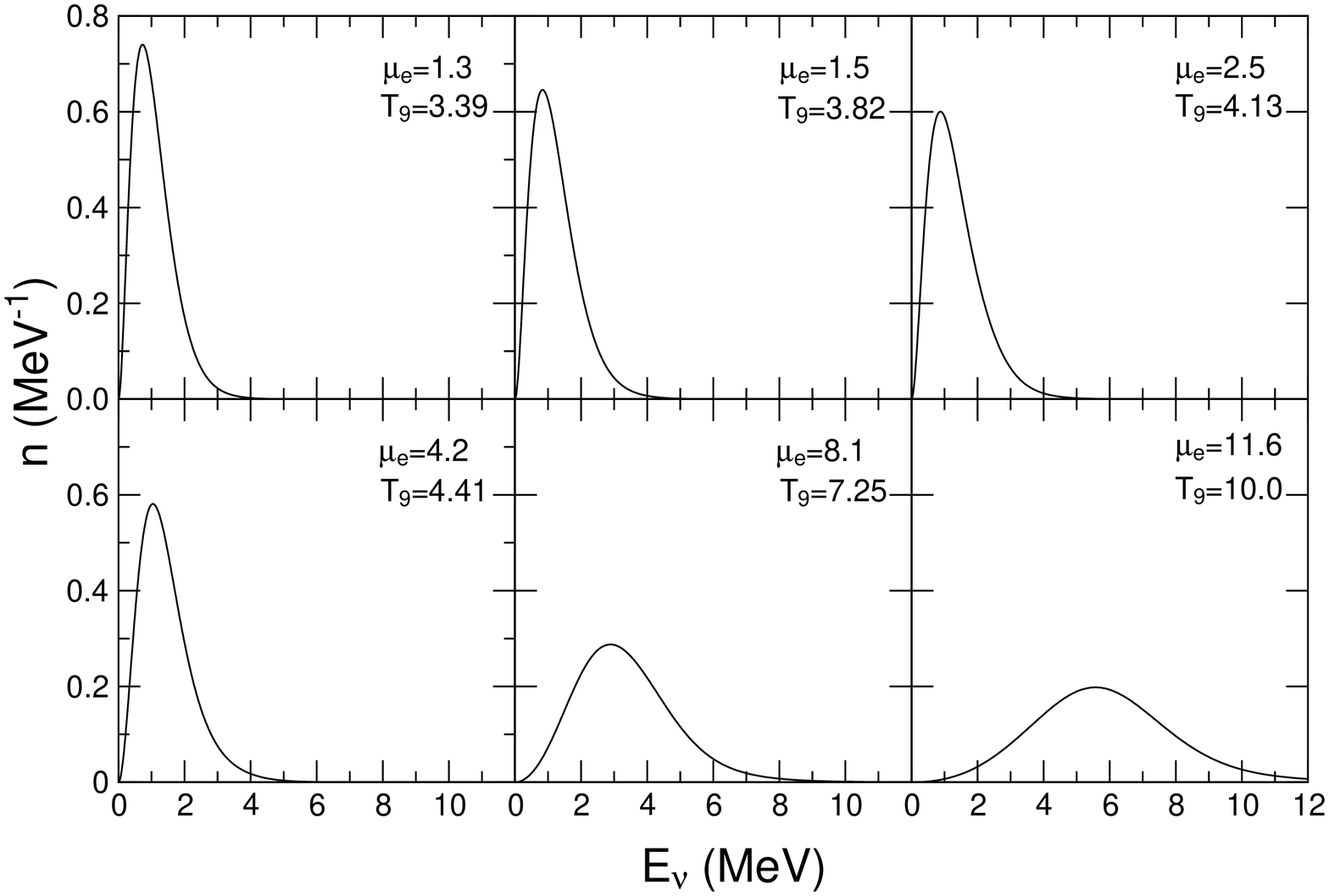}
  \end{center}
  \caption{Normalized neutrino spectra for stellar electron capture on 
    $^{56}$Fe at several different phases of the final evolution of a
    15~M$_\odot$ star. The stellar parameters have been taken from
    Table~1 of \protect\cite{Heger00}. The last panel corresponds to
    typical conditions during the collapse phase ($T=10^{10}$ K,
    $\rho=3 \times 10^{10}$ g/cm$^3$, and $Y_e=0.42$). The chemical
    potentials are in MeV, while $T_9$ defines the temperature in
    $10^9$ K.}
\end{figure}

As examplified for $^{56}$Fe in Fig.~4 the situation is quite
different if one studies the neutrino spectrum emerging from capture
on an even-even nucleus as function of electron chemical potential. To
understand the reason we note two facts. First, as the GT distribution
in the daughter is quite concentrated at low excitation energies a
double-peak structure related to the distinct capture to low-lying
states and the GT bulk does not emerge.  Second, in even-even nuclei
the backresonances are at higher excitation energies than in odd-odd
nuclei \cite{Langanke00}.  The consequences for the neutrino spectrum
are obvious.  If compared to $^{56}$Co, the relative contribution of
the low-lying states to the electron capture on $^{56}$Fe is
significantly enhanced with respect to the backresonances and no
high-energy neutrino peak emerges.  As an increase of the electron
chemical potential makes the capture energetically easier, the
neutrino peak energy moves to higher energies.  We mention that the
last panel of the figure already corresponds to a phase of the
contraction after the presupernova model which we have approximated by
$T=10^{10}$ K, $\rho=3 \times 10^{10}$ g/cm$^3$ and $Y_e=0.42$.  It
demonstrates that the simple structure of the spectrum remains also
during that stellar evolution stage which requires detailed neutrino
transport.

As stated above detailed neutrino transport becomes important in the
final evolution of massive stars, following the presupernova models.
Due to electron captures the matter in the final presupernova models
is neutronrich and the nuclei present have negative $Q_0$-values. On
the other hand, the electron chemical potential has grown strongly
enough until this point thus allowing capture to the bulk of the GT
strength. Fig. 5 shows the neutrino spectra for the 6 nuclei which
dominate electron capture in the presupernova models of a 15~M$_\odot$
star \cite{Heger00}.  Due to Heger {\it et al.} the core density and
temperature are $\rho=9.1 \times 10^9$ g/cm$^3$ and $T=7.2 \times
10^9$ K, while the $Y_e$ value is 0.432 \cite{Heger00}.  The chemical
potential then is $\mu_e=8.1$ MeV. We note that all neutrino spectra
are single-peaked. The average neutrino energy released by nuclei is
about 3 MeV, while it is 6.25 MeV for capture on free protons which
under these presupernova conditions become abundant enough to
significantly contribute to electron capture.

\begin{figure}[thb]
  \begin{center}
    \includegraphics[width=0.9\columnwidth]{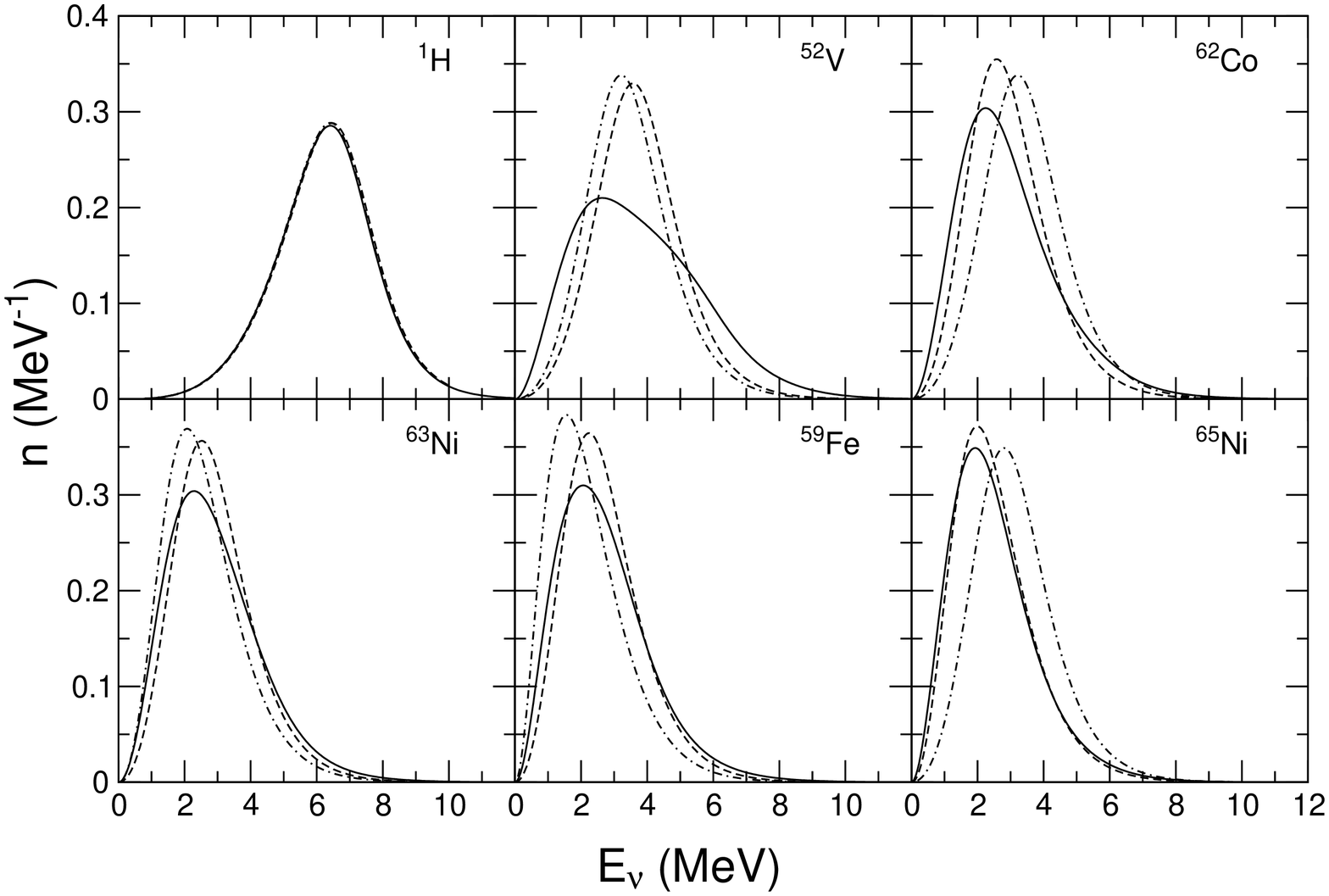}
  \end{center}
  \caption{Normalized neutrino spectra for stellar electron capture on
    the six most important `electron-capturing nuclei' in the
    presupernova model of a 15 M$_\odot$ star, as identified in
    \protect\cite{Heger00}.  The stellar parameters are $T=7.2 \times
    10^9$ K, $\rho=9.1 \times 10^9$ g/cm$^3$, and $Y_e=0.43$. The
    solid lines represent the spectra derived from the shell model
    electron capture rates. The dashed line shows the fit to the
    spectra, using the parametrization of Eq. (6) and adjusting the
    parameter $q$ to the average neutrino energy of the shell model
    spectrum. The dashed-dotted spectrum corresponds to the
    parametrization recommended in \protect\cite{Bruenn85}.  }
\end{figure}

As neutrino cross sections scale with $E_\nu^2$, high-energy neutrinos
are more easily trapped.  Capture on free protons has a more favorable
$Q_0$ value than capture on neutronrich nuclei present in the
presupernova matter composition.  As a consequence the neutrino
average energy is higher for capture on free protons. Nevertheless
captures on nuclei still produces the larger amount of high-energy
neutrinos in lighter stars (M $<$ 20 M$_\odot$) as they dominate the
rate in the presupernova models. This is different for heavier stars
where the capture on free protons in the presupernova model
corresponds already to 30--50~\% \cite{Heger00}.

\begin{figure}[thb]
  \begin{center}
     \includegraphics[width=0.9\columnwidth]{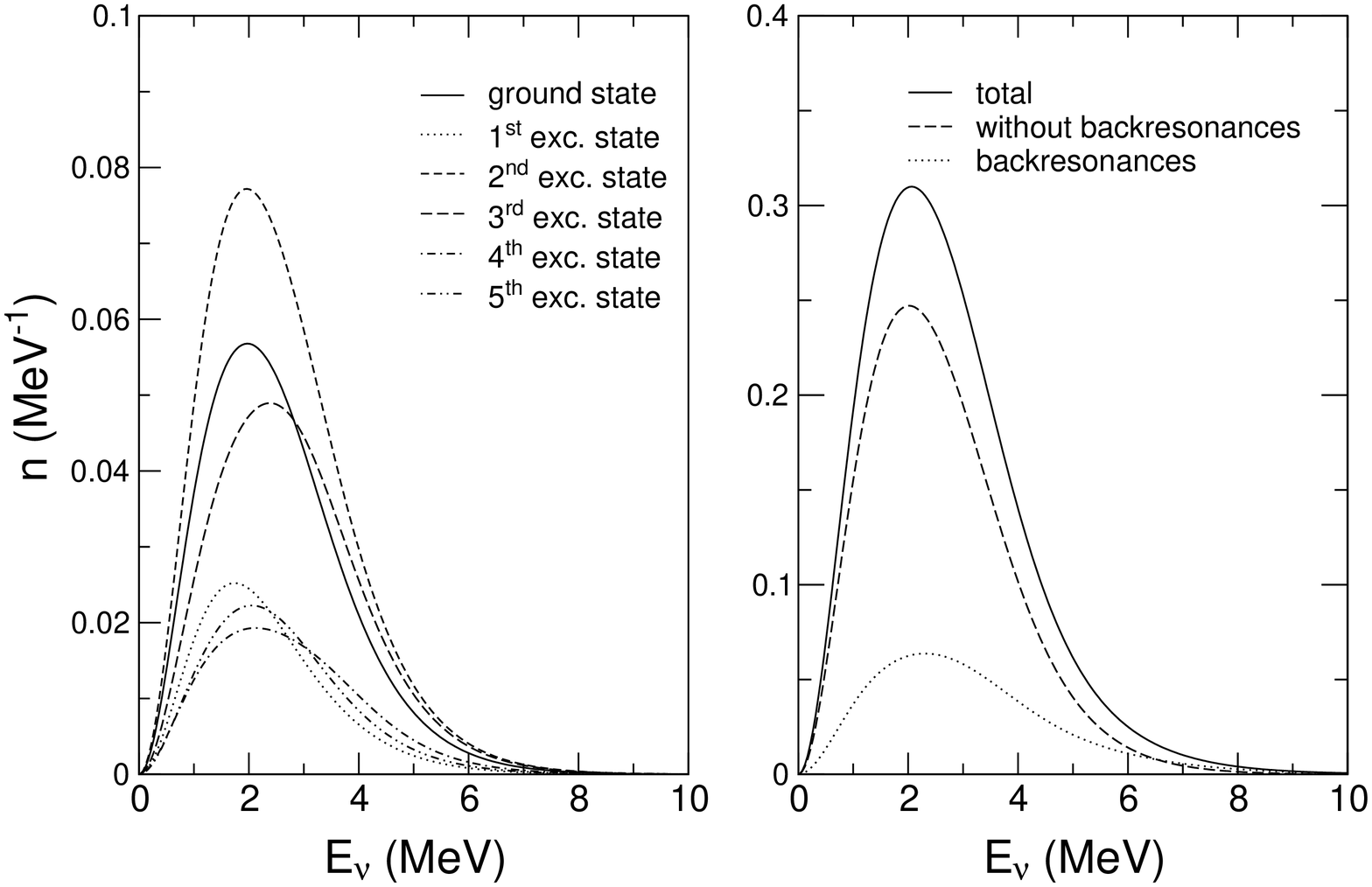}
  \end{center}
  \caption{Partial contributions of individual states in the parent
    nucleus to the Neutrino spectrum for stellar electron capture on
    $^{59}$Fe. The calculation has been performed for the presupernova
    conditions of Fig. 5. The spectra are multiplied by their relative
    weight to the total capture rate. The left panel shows the
    neutrino spectra calculated for the ground state (solid line) and
    the 5 lowest excited states in $^{59}$Fe. The right panel compares
    the total neutrino spectrum (solid line) with the one obtained
    only from the backresonances (dotted line, for a definition see
    text and \protect\cite{Langanke00}). The dashed line shows the
    contributions from the low-lying individual states which have been
    considered to calculate the electron capture rate
    \protect\cite{Langanke00}.  }
\end{figure}

For presupernova conditions the shell model evaluation predicts
slightly larger average neutrino energies than the FFN rates
\cite{Heger00}.  A possible explanation for this difference is given
by the fact that the shell model rates explicitly consider capture
from thermally excited states.  The Q-value for $^{59}$Fe is
$Q_0=-5.696$ MeV. However, the excited states have more favorable
$Q_{\text{eff}}$ values than the ground state thus making electron
capture more easy and in several cases supporting larger neutrino
energies.  Fig. 6 shows the neutrino spectrum calculated for the 6
lowest states in $^{59}$Fe. (Here the distributions are multiplied by
their relative weight in the total rate).  One finds that indeed
neutrino spectra from excited states often have a wider tail. This
results from capture to low-lying states in the daughter. In contrast
to the low-lying transitions the GT bulk approximately obeys Brink's
hypothesis, e.g.  \cite{Brink}.  This states that the GT strength of
excited states is the same as for the ground state, only shifted by
the excitation energy of the parent state. In particular, Brink's
hypothesis implies that the relevant energy difference
$Q_{\text{eff}}-E_j$ is the same for capture to the GT bulk for all
parent states.  Another observation has already been mentioned above.
Although the total capture rate is still dominated by the transition
to low-lying states, the main source for high-energy neutrinos,
however, are the `backresonances' \cite{FFN1,FFN2}.

If the shell model neutrino spectra are to be used in collapse
simulations, they have to be represented by parametrizations which are
accurate, fast and can be easily implemented. Our proposal for the
parametrization is based on the following approximations. Suppose that
i) the electron capture on a state in the parent nucleus, described by
GT transitions, leads to a single state in the daughter nucleus at
energy $E_f^*$ and that ii) Brink's hypothesis is valid, i.e. this
state is at $E_f^*+E_i$ if the capture is on an excited state in the
parent at excitation energy $E_i$.  If we equal electron energy and
momentum, which is a valid approximation for the conditions we are
interested in, the neutrino spectrum has the form \cite{Bruenn85}
\begin{equation}
n(E_\nu) = E_\nu^2 (E_\nu-q)^2 \frac{N}{1 + \exp \left\{ (E_\nu-q -
    \mu_e)/kT 
\right\}}
\end{equation}
with $q=Q_0 - E_f^*$ and a constant $N$ which normalizes the neutrino
spectrum to unity. This form is obviously valid for capture on free
protons where the parameter $q$ is the reaction $Q_0$ value, $q=-1.29$
MeV. For finite nuclei, $q$ should be considered a fit parameter. It
can be adjusted to the average neutrino energy which is listed in the
shell model rate tabulations \cite{ADNDT} for a grid of
temperature/density points and can be easily interpolated in-between.
We have tested this proposal and generally find that the
parametrization approximates the shell model spectra rather well, as
can be seen by the dashed curves in Fig. 5.  Of course, our
parametrization fails if the spectrum is double-peaked as observed for
capture on odd-odd nuclei under special conditions (e.g. see Fig. 1.).
However, we do not expect that these conditions occur for $pf$-shell
nuclei during the collapse phase where the electron chemical potential
is high enough compared to the reaction $Q_0$ value to allow
appreciable electron capture to the GT bulk. Finally we note that for
$^{52}$V the shell model spectrum is wider than the parametrization.
This is caused by the fact that the spectrum for the individual states
shows noticeable differences and does not strictly follow the Brink
hypothesis. For example, the excited state of $^{52}$V at 22 keV has
$J=5$ and thus there are no low-lying states in the daughter $^{52}$Ti
which can be reached by GT transitions. This is different for the
excited $J=1$ state at 141 keV which connects strongly to the
$^{52}$Ti ground state.

Bruenn suggested a similar parametrization for the neutrino spectra
emerging from electron capture on $pf$-shell nuclei, however simply
setting $q=Q_0-3$ MeV \cite{Bruenn85}.  The resulting spectra are
compared to the shell model spectra in Fig. 5. Despite the simple
guess for the parameter, the agreement is quite acceptable.

In summary, the knowledge of the neutrino energy spectra at every
point and time in the core is quite relevant for simulations of the
final collapse and explosion phase of a massive star. In the collapse
phase, neutrinos are mainly produced by electron capture on nuclei and
protons and their emerging energy spectra are an important ingredient
in the simulations. In this paper we have presented neutrino spectra
for stellar electron capture during the final presupernova evolution
stage of massive stars.  The spectra have been consistently derived in
the framework of the recently evaluated capture rates which have been
calculated on the basis of state-of-the-art large-scale shell model
studies. Furthermore we have calculated the spectra for stellar
conditions which have been obtained in presupernova evolution of
massive stars, using the same shell-model weak interaction rates. The
calculated presupernova neutrino spectra show a rather simple
structure which is easily parametrizable.  This parametrization is
easily implementable into the simulation codes and allows for a
derivation of the neutrino spectra consistent with the shell-model
weak-interaction rates.
 
\acknowledgements 

Our work has been supported by the Danish Research Council. GMP thanks
the Carlsberg Foundation for a fellowship. JMS acknowledges a
scholarship of the Funda\c{c}\~ao para a Ci\^encia e Tecnologia.

\end{document}